# Do Gravitational Fields Have Mass ?
## Or on the Nature of Dark Matter

Ernst Karl Kunst

**As has been shown before (a brief comment will be given in the text), relativistic mass and relativistic time dilation of moving bodies are equivalent as well as time and mass in the rest frame. This implies that the time dilation due to the gravitational field is combined with inertial and gravitational mass as well and permits the computation of the gravitational action of the vacuum constituting the gravitational field in any distance from the source of the field. Theoretical predictions are compared with experimental results and it is shown that many known astrophysical and gravitational phenomena, especially the so-called dark or missing matter, owe their existence to the gravitational effects of the mass of the field-vacuum.**

## Introduction

Apparent deviations from the Einstein-Newtonian law of gravitation both on laboratory and astronomical scale have been known long since. Those partly controversially discussed gravitational phenomena are:

1) Constantly high velocities of individual galaxies within clusters and groups of galaxies, departing strongly from the velocities on the strength of the virial law and constantly high orbital velocities in the vicinity of the Milky Way, other galaxies and galaxy pairs, which deviate strongly from a Keplerian velocity distribution. Both phenomena have led to the currently accepted concept of non-luminous, non-baryonic material in the vicinity of large systems on a cosmic scale, the so-called "halo of dark unseen matter" [1], [2];

2) An apparent increase of the universal gravitational constant G with growing radial distance of test masses measured with the torsion pendulum in the laboratory [3];

3) A systematic increase of the gravitational acceleration g as one descends into deep mineshafts or boreholes [4], [5], [6], or decrease as one ascends towers [7];

4) A systematic linear deviation of the acceleration of two test masses at the ends of the torsion pendulum in the gravitational field of Earth in proportion to the difference in baryon density (protons plus neutrons per unit mass), which was found by analytical replication of the original Eötvös data and led to the suggestion of a composition-dependent finite range repulsive ( fifth ) force [8];

5) A systematic decrease of the velocity of space-probes on their track

outbound of the solar system as e. g. Pioneer 10 and 11 [9].

In the following we will show that all these experimentally found though - as already stated - partly controversially discussed phenomena are due to the gravitational effects of the mass of the gravitational field.

### Connection between Relativistic Mass and Dilated Time of Moving Bodies

Main results of the modified theory of relativistic kinematics among others are inertial motion (velocity) always to be symmetrically composite and the Lorentz transformation not to predict the Fitzgerald-Lorentz contraction of the dimension ($\Delta x$) parallel to the velocity vector, as invented by Fitzgerald and Lorentz to account for the null-result of the Michelson-Morley experiment on moving Earth, but rather an expansion of $\Delta x$ [10]. Accordingly the volume **V'** of an inertially moving body will any observer resting in a frame considered at rest seem enhanced by the factor

$$V' = \Delta x' \Delta y' \Delta z' = \Delta x \gamma_0 \Delta y \Delta z = V\gamma_0, \quad (1)$$

where **V** means volume and $\gamma_0$ the Lorentz factor based on the composite velocity $v_0$. Among others it has been demonstrated this expansion of $\Delta x$ (or **V**) to be the cause of the experimentally found increase of the interaction radius respectively cross section of elementary particles with rising energy (velocity), as determined in collider experiments and as is known from studies of cosmic radiation. From m' = **V'**ρ' = mγ$_0$ = **V**ργ$_0$ in connection with (1) follows ρ' = ρ and, therewith, the fraction $v_0/c$ of the relativistically dilated time to be the very cause of the relativistic increase of mass:

$$dt' c v_0 = E_t = m_t c^2,$$
$$m_t = \frac{E_t}{c^2} = \frac{dt' v_0}{c} = \frac{dx'}{c}, \quad (2)$$

where "ρ" means density of mass, "$E_t$" energy of the product dt'cv$_0$ of a moving material body and "$m_t$" mass induced by time dilation. Furthermore has been demonstrated mass of the hydrogen (H-) atom and quantum of time $2\lambda_1/c$ in the rest frame be equivalent and generated by the movement of a fourth spatial dimension of the atom

$$\frac{2\lambda_1}{c} = \frac{\sqrt{h}}{c} = m,$$

where "$2\lambda_1$" is the fundamental length in $R^4$, "h" Planck's constant and "m" mass of the smallest elctrically neutral and stable piece of matter, presumedly of the H-atom [11]. Analogous to the equivalence of mass and time in the rest frame as well as in the moving one gravitational fields have to be considered to be spaces with relativistic mass, because of the time dilation due to the gravitational field. Thus, we can expect the potential differences in gravitational fields in dependence on the distance from the source of the field to be perceptible as physically measurable masses. To compute those masses basically two possibilities exist.

## Global Estimate of the Mass of the Gravitational Field

We refer to the classical definition of mass as the product of volume and density and then in accord with (2) can write

$$E_V = m_V c^2 = dt' c v_0 \times c\,dy\,dz = V' c v_0,$$

wherefrom follows

$$E_V = m_V c^2 = V c v_0 \tag{3}$$

if $v_0 \ll c$ and $dt/dt' \approx 1$, where $E_V$ means energy and $m_V$ mass of volume of space (vacuum), respectively. Of course, (3) is valid in the case of moving bodies only, with one velocity vector.

Suppose we have a spherically symmetric gravitational field in the form of a Schwarzschild-vacuole in the Friedmann cosmos

$$r^3 = 6 \frac{M}{\varkappa \rho c^2} = K^3 \varrho^3,$$

where "r" means the coordinate radius of the vacuole and "$\varrho$" the "radius", "K" means curvature, "M" the mass of the central body under consideration [12]. In this case the product of each of the three dimensions of the vacuole and the velocity vector or (in Newtonian approximation) scalar of curvature

$$K = \frac{r}{\varrho} \approx \frac{v_0}{c} \approx \frac{2GM}{c^2 R_1}$$

will contribute to the global energy of the vacuole so that according to (3) the energy content of the space of the vacuole - as seen from "outside" - can be written as

$$M_V c^2 = V_{\varrho_x \varrho_y \varrho_z} c K^3 = \varrho^3 c K^3 = \varrho^3 c \left( \frac{2GM}{c^2 R_1} \right)^3, \tag{4}$$

whereby "G" is the Newton's gravitational constant and "$R_1$" the radius of the mass of the central body, distributed in the vacuole. Here we had to consider that each of the three geometric dimensions of the gravitational field must be multiplied by the scalar of curvature or (in approximation) vector of velocity, according to the principle of equivalence. On the grounds of the ratio

$$\rho_0 = \frac{\text{mass in the cosmos}}{\text{volume of the cosmos}} \approx \frac{\text{mass (of the body) in the vacuole}}{\text{volume of the field (vacuole)}}$$

- whereby $\rho_0$ means cosmic density of mass - we eventually can approximate the total mass of the gravitational field of the vacuole with the expression

$$M_V = \frac{\varrho^3}{c}\left(\frac{2GM}{c^2 R_1}\right)^3 \approx \frac{M}{\rho_0 c}\left(\frac{2GM}{c^2 R_1}\right)^3 .$$

Accordingly the ratio of the "total mass of the gravitational system" (field + visible mass) to the "visible mass" is

$$\frac{M_{total}}{M} = \frac{M + M_V}{M} = 1 + \frac{1}{\rho_0 c}\left(\frac{2GM}{c^2 R}\right)^3 .$$

Applying the corresponding values for galaxies and galaxy clusters, roughly the right amount of masses results i.e. greater than the visible masses which were introduced by astronomers as the so called "dark or missing matter", to explain the dynamics of large complexes of gravitational systems. The uncertainties in the determination of the cosmic density $\rho_0$ and especially of $R_1$ (boundary between the external and internal Schwarzschild metric) in extended gravitational systems e.g. galaxies and clusters of galaxies allow but only very global estimates.

### The Mass of the Gravitational Field in Dependence on the Distance from the Source of Gravitation

More exact results can be derived if, starting at the boundary between the external and internal Schwarzschild metric of a spherical distribution of mass (source of the external field or radius of the mass in Euclidian coordinates), the infinitesimal small distances respectively multiplied by the time dilation at the point of the coordinate in radial direction are summed. Consider the center of gravity of the field-producing mass to be at rest, with $T_{44}$ being the essential component of the energy momentum tensor $T_{ik}$ so that in a first approximation is valid

$$g_{44} = -\left(1 + \frac{2GM}{c^2}\right), \quad GM = -\frac{\varkappa c^2}{8\pi}\int \frac{T_{44} d^3 x}{|x^\alpha - x^{-\alpha}|} .$$

Furthermore, the vacuum of the external Schwarzschild metric is considered to consist of thin concentric shells of the thickness dR, where "R" is the distance from the center of gravity. According to (4) the mass of each infinitesimal thin concentric shell measured from the point $R_1$ radially "within" the field must be proportional to the infinitesimal small distance dR:

$$dM_{v_R} = \frac{v_0 \, dR}{\sqrt[3]{c}} = \text{constant} .$$

The volume of each successive shell increases as the square of the radius. On the other hand the velocity $v_0$ decreases inversely proportional to the radius, i.e. the cube of the velocity $v_0^3$ with increasing radius, as $1/R^2$. As a result, the product of each successive shell by the cube of the respective velocity remains constant. Thus, the mass of the vacuum of each successive shell of the gravitational field remains constant for all R. Because

$$v_0 = v_{0_R} \sqrt{R} = \text{constant}$$

for all R - where $v_{0R}$ means velocity at the point R - and

$$v_{0_R} = \left(1 - \frac{2GM}{R}\right)^{-\frac{1}{2}} - 1 \approx \frac{GM}{R} ,$$

the proportion of mass of the infinitesimal part dR of the radius R of the gravitational field amounts to

$$dM_{v_R} = \frac{v_{0_R} \sqrt{R} \, dR}{\sqrt[3]{c}} = \frac{GM \, dR}{\sqrt[3]{c} \sqrt{R}} .$$

Integration results in the mass of the field vacuum within the radius R

$$M_{v_R} = \int_{R_1}^{R} \frac{GM \, dR}{\sqrt[3]{c} \sqrt{R}} = \frac{2GM(\sqrt{R} - \sqrt{R_1})}{\sqrt[3]{c}} + \Lambda , \qquad (5)$$

measured within the field. $R_1$ means the radius of the internal Schwarzschild metric ("radius" of the mass), R the radial distance from the center of gravity of the field producing mass, measured in Euclidean coordinates. The constant $\Lambda$ is the mass of the macroscopic groundstate of vacuo (not disturbed by gravitational fields), which is null, and $M_{vR}$ the mass of the vacuum of the gravitational field in the distance R from the center of gravity which together with the field-producing mass M or the energy momentum tensor $T_{ik}$, respectively, determines completely the behaviour of test bodies of the mass m.

In principle this result is also valid for the internal Schwarzschild metric, because the space inside a gravitational body contributes to the total mass of the body or density of energy $T_{ik}$ in the distance $R_1$ from the center of gravity (in approximation):

$$M_{V_{R_1}} = \frac{2GM\sqrt{R_1}}{\sqrt[3]{c}}.$$

Thus, if $R < R_1$ the mass of the gravitational field inside the mass also amounts to the value (5), measured from the point $R_1$ in negative radial direction or toward the center of gravity.

## Comparison of Theoretical Predictions with Experiment

1) According to (5) the total mass of a gravitational system in the radial distance R from the center of gravity of the field producing mass M amounts to

$$M_{total} = M + M_{V_R} = M + \frac{2GM(\sqrt{R}-\sqrt{R_1})}{\sqrt[3]{c}}, \qquad (6)$$

or after division by M the quotient is given by

$$\frac{M_{total}}{M} = \frac{M + M_{V_R}}{M} = 1 + \frac{2G(\sqrt{R}-\sqrt{R_1})}{\sqrt[3]{c}}. \qquad (7)$$

From (6) the orbital Kepler velocity of a body of negligible mass as a function of R, $R_1$ and the central mass M is derived:

$$v_{0_R} = \sqrt{\frac{GM}{R}\left(1 + \frac{2G(\sqrt{R}-\sqrt{R_1})}{\sqrt[3]{c}}\right)}. \qquad (8)$$

Computation results in the flat non-Keplerian rotation curves of galaxies and pairs of galaxies established by astronomical observations, whereby the morphology of the curve strongly depends on $R_1$. Calculation of (7) results directly in the ratio of the total perceptible mass within the distance R of the gravitational field - baryonic plus field - to the amount of the luminous matter, which agrees well with astronomical measurements. In the following we compare theoretically derived values of $M_{total}/M$ according to (7) with some experimental results for the outer regions of the Milky Way [1]:

| $M_{total}/M$ (7) | Experiment | R (kpc) | $R_1$ (kpc) | "m" |
|---|---|---|---|---|
| 2.6 | ≈3.0 | 18 | 10 | carbon monoxide clouds |
| 11.0 | ≈9.0 | 60 | 10 | clouds of Magellan |
| 13.6 | ≈12.0 | 75 | 10 | satellite galaxies |

$R_1$ = 10 kpc is the mean distance of the sun from the galactic center, because astronomical measurements are grounded on the validity of the Newton-Keplerian law within the orbit of the sun (see also 5) below).

Equation (7) results also convincingly in the linear increase of $M_{total}/M$ with growing R in vast cosmic systems as measured by astronomers [1], [2]:

| $M_{total}/M$ (7) | Experiment | R | $R_1$ | "M" |
|---|---|---|---|---|
| 10 | ≈10 | 100 kpc | 0 | galaxies |
| 25 | ≈25 | 100 kpc | 0 | pairs of galaxies |
| 430 | >400 | 32 mpc | 0 | Coma cluster |
| 650 | >600 | 70 mpc | 0 | local supercluster |

2) From the preceding is evident that measurements of the R-dependence of the acceleration in local fields of gravitation must yield apparent discrepancies to Einstein-Newtonian gravity, which usually are interpreted either as a modification of the gravitational constant G or as the effect of an additional (fifth) force of nature. A direct measurement of the gravitational force f, which a unit of mass M = 1 exerts on a test mass m in the distance R from the center of gravity results according to (5) in an additional acceleration:

$$\Delta a = \frac{\Delta f}{m} = \Delta G = \frac{2G(\sqrt{R} - \sqrt{R_1})}{\sqrt[3]{c}} ,$$

$$G_R = G + \Delta G = G\left(1 + \frac{2(\sqrt{R} - \sqrt{R_1})}{\sqrt[3]{c}}\right) . \qquad (9)$$

Evidently $G_R$ expresses an apparent alteration of G due to the gravitational effect of the field vacuum.

In 1976 Long compared older measurements at various ranges of R with the results of his own torsion pendulum experiments at R = 4.5 cm and 30 cm and found

$$G_R \approx G[1 + 0.002 \ln(R)] \qquad (9a)$$

on laboratory scale [3]. For an overview we compare theoretical and experimental results :

| Theory<br>ΔG = $G_R$ - G (9) | Experiment<br>ΔG ≈ $G_R$ - G (9a) | R<br>cm | $R_1$ |
|---|---|---|---|
| 0.0135 | 0.0077 | 10 | 0 |
| 0.0235 | 0.0223 | 30 | 0 |
| 0.0303 | 0.0291 | 50 | 0 |
| 0.0358 | 0.0336 | 70 | 0 |

where G = 6.656 × $10^{-8}$ $g^{-1}$ $s^{-2}$ according to Long.

3) The influence of the mass of vacuo constituting the gravitational field of Earth on the gravitational acceleration g results according to (8) in:

$$\Delta g = \sqrt{\frac{2G^2 M_E(\sqrt{R_E} - \sqrt{R_1})}{\sqrt[3]{c}\, R_E}},$$

$$G_{(M_E + M_v)} = G\left(1 + \frac{R_E^2}{GM_E}\sqrt{\frac{2G^2 M_E(\sqrt{R_E} - \sqrt{R_E - R_1})}{\sqrt[3]{c}\, R_E}}\right)$$

$$= G\left(1 + R_E\sqrt{\frac{2R_E(\sqrt{R_E} - \sqrt{R_E - R_1})}{\sqrt[3]{c}\, M_E}}\right), \quad (10)$$

or as an apparent alteration of the gravitational constant of the amount $G_{(ME + MV)}$ = G + ΔG, where $M_E$ means mass and $R_E$ radius of Earth, respectively - $R_1$ is here the negative radial direction toward the center of Earth measured from the point $R_E$.

Consistently higher values of G from measurements of g in boreholes and mines for some time have been known to point to a deviation from the 1/R-law of the gravitational potential. Therefore, a direct comparison of this theory with experimental results is possible. In the following we compare some results of Stacey [4], Holding [5] and Hsui [6] from measurements in boreholes and mines with calculations according to (10):

| Theory<br>$G_{(ME + MV)}$ (10) | Experiment<br>$10^{-8}$ $cm^3$ $g^{-1}$ $s^{-2}$ | $R_1$<br>cm |
|---|---|---|
| 6.674 | 6.724 ± 0.014 | 2 × $10^4$ |

| 6.722 | 6.734 ± 0.002 | 1 × 10⁵ |
| 6.727 | 6.700 ± 0.065 | 1.2 × 10⁵ |
| 6.772 | 6.810 ± 0.070 | 4 × 10⁵ |

where $G = 6.672 \times 10^{-8}$ cm$^3$ g$^{-1}$ s$^{-2}$, $M = 5.97 \times 10^{27}$ g, $R_E = 6.4 \times 10^8$ cm.
Eckhardt measured in a tower experiment at $R_1 \approx 6 \times 10^4$ cm above the ground a deviation of g of 500 ± 35 µGal [7], whereas our formula delivers 400 µGal, where $g = 981$ cm s$^{-2}$.
Of course, in the case of the tower experiment $(R_E - R_1)^{1/2}$ in (10) must be replaced by $(R_E + R_1)^{1/2}$. The coincidence of theory and experiment does not look very impressive - which easily is explained by the tremendous uncertainties on the experimental side -, but nevertheless, a systematic trend clearly shows up.

4) A look at (10) shows that the apparent alteration of G due to the mass of the gravitational field should not only be dependent on the distance R, but also on the composition of the material of the attracted mass. The reason is that if R = constant, then the fraction $R_1/R$, and thereby the difference $(\sqrt{R} - \sqrt{R_1})$, differs with the density ρ of the material. The density ρ is the determining parameter and it is clear that ρ is appproximately related to the difference in baryonic density. In other words, we assume the mass of the field vacuum also to play a passive role as attracted mass. For convenience we choose

$$\rho_1 = 1, \quad R_{\rho_1} = \sqrt[3]{\rho_1}, \quad \sqrt{R} - \sqrt{R_{\rho_1}} = 0,$$

constituting the density of H$_2$O as the reference value. Because stock density varies inversely to the volume per unit mass, relative to H$_2$O the radius $R_1$ of the unit mass of all materials other than H$_2$O varies as $1/\sqrt[3]{\rho}$ so that (5) attains the form:

$$M_{V_{(R_1-R)}} = \frac{2GM}{\sqrt[3]{c}} \left(1 - \frac{1}{\sqrt[6]{\rho}}\right).$$

As compared with the mass of the gravitational field of an unit mass H$_2$O, which we arbitrarily set zero, the mass of the field of the unit mass of a material other than H$_2$O varies as $(1 - 1/\sqrt[6]{\rho})$. Correspondingly, the gravitational force acting upon a test mass in a locally (almost) homogeneous field, particularly that of Earth, must vary proportional to the density of the test body as:

$$f = g(M + M_{V_{(R_1-R)}}) = M(g + \Delta g) = gM\left[1 + \frac{2G}{\sqrt[3]{c}}\left(1 - \frac{1}{\sqrt[6]{\rho}}\right)\right], \tag{11}$$

which means: Bodies of equal mass but different density (baryonic density) experience an apparent composition - dependent relative gravitational acceleration, which is due to small differences of the integrated masses of the respective gravitational fields.

Fischbach's analysis of the old Eötvös data includes among others three pairs of sample material: $H_2O$-Cu, asbestos-Cu and Pt-Cu [8]. The experimental results for these pairs are very convenient for a comparison with theory. Computation of (11) and comparison with the results of Fischbach-Eötvös in g(g) × $10^9$ results in (experiment in brackets):

$H_2O$ - asbestos =   6.67 ($\approx 7 \pm 2$)
$H_2O$ - Pt       = 16.93 ($\approx 14 \pm 2$),
whereby $\rho_{asbestos}$ = 2.8 and $\rho_{Pt}$ = 21.45; $\rho$ means density.

5) The mass of the field vacuum surrounding the sun amounts in any distance R (from the sun) acording to (5) to

$$M_{V_R} = \frac{2 G M_\odot (\sqrt{R} - \sqrt{R_1})}{\sqrt[3]{c}}.$$

Thus, the gravitational pull of the mass $M_V$ in the distance R from the sun must be

$$\frac{G M_{V_R}}{R^2} = \frac{2 G^2 M_\odot (\sqrt{R} - \sqrt{R_1})}{\sqrt[3]{c}\, R^2}. \qquad (12)$$

Pioneer 10 is currently 71 times as far from the sun as Earth is. According to (12) the gravitational pull of the field vacuum of the sun in this distance R ≈ 1.06216 × $10^{15}$ cm onto the spacecraft must be $GM_V/R^2$ = 10.51 × $10^{-8}$ cm $s^{-2}$ - where $M_\odot$ = 2 × $10^{33}$ g and $R_1$ = 1.428 × $10^{14}$ cm, the mean distance of Saturn from the sun -, whereas Anderson reported an experimentally found acceleration of ≈ 8.5 × $10^{-8}$ cm $s^{-2}$ toward the sun [9]. To choose the proper value of $R_1$ in (12) it had to be considered that analogous to the case of the Milky Way astronomical measurements are grounded on the validity of the Newton-Keplerian law within the orbit of Saturn, or with other words: in all computations on the grounds of Einstein-Newtonian gravitation the mass of the field vacuum of the sun (and of the planets) is at least till the orbit of Saturn included in the mass of the sun. If $R_1$ = 2.872 × $10^{14}$ cm - the mean distance of Uranus - (12) yields 7.97 × $10^{-8}$ cm $s^{-2}$. Besides we have to expect that the straightforward application of (12) to the gravitational field of the sun is restricted for the following reasons:
If the mean distances between the planets are listed in A. U. according to the Titius-Bode law (which with the exception of Pluto correspond roughly to the observed distances) the following ratios result:

| | | | | | |
|---|---|---|---|---|---|
| Mercury - Venus   | : Venus   - Earth   | =  0.3 :  0.3 | = 1 : 1, |
| Mercury - Earth   | : Earth   - Mars    | =  0.6 :  0.6 | = 1 : 1, |
| Mercury - Mars    | : Mars    - Ast     | =  1.2 :  1.2 | = 1 : 1, |
| Mercury - Ast     | : Ast     - Jupiter | =  2.4 :  2.4 | = 1 : 1, |
| Mercury - Jupiter | : Jupiter - Saturn  | =  4.8 :  4.8 | = 1 : 1, |
| Mercury - Saturn  | : Saturn  - Uranus  | =  9.6 :  9.6 | = 1 : 1, |
| Mercury - Uranus  | : Uranus  - Neptune | = 19.2 : 19.2 | = 1 : 1, |
| Mercury - Neptune | : Neptune - Pluto   | = 38.4 : 38.4 | = 1 : 1. |

Because - as shown before - the mass of each successive shell (being proportional to the distances between the planets or the "thickness" of the shells) of the gravitational field of the sun remains constant, the above ratios seem to indicate that in the protoplanetary disk and later the planets positioned more or less exactly between field shells of equal mass. This can be described as

$$r_{n+1} = 2^{n-2}(1 - r_1) + r_1 , \qquad (13)$$

where n = 1, 2, 3,...n and $r_3$ = 1. Inserting $r_1$ = 0.4 in (13) delivers again the Titius-Bode law $2^{n-2}$ × 0.3 + 0.4. It s clear that all r > $r_1$ depend on the value of $r_1$, which again cannot be derived from (13). Obviously is the simple rule of balance of field mass shells, developed above, not straightforwardly applicable to the three innermost solar planets. But if our hypothesis is correct, must their distances from the sun also depend on the balance of the field masses. Hence they should tend to take positions between three shells of equal mass at (in arbitrary units) $r_1$ = 0.33, $r_2$ = 0.66 and $r_3$ = 1. On the other hand there must exist a tendency to form three shells at $r_1$ = 0.5, $r_2$ = 0.75 and again $r_3$ = 1 to reach a balance 0.5 : 0.5. As a consequence the planets tend to take position between $r_1$ = 0.33 and 0.5, which results in $r_1$ = 0.41, and $r_2$ = 0.66 respectively 0.75, which results in $r_2$ = 0.71.
Thus, if this hypothesis is correct, it must be valid for any system, where at least three objects (with a similiar genesis as the planets of the sun) orbit a central mass. If always $r_3$ = 1 (in the case of Saturn the mean distance of Tethys to Calypso) we find for the two innermost objects e. g. in the system of the sun $r_1$ = 0.39 and $r_2$ = 0.72, Jupiter $r_1$ = 0.39 and $r_2$ = 0.63, Saturn $r_1$ = 0.49 (mean distance of Atlas to Epimetheus) and $r_2$ = 0.72 (mean distance of Mimas and Enceladus), Uranus $r_1$ = 0.48 and $r_2$ = 0.72, the pulsar PSR 1257 + 12 $r_1$ = 0.4 and $r_2$ = 0.77, and the pulsar PSR 1828 - 11 $r_1$ = 0.44, and $r_2$ = 0.63, respectively.